\documentclass[
]{ceurart}

\usepackage{listings}
\usepackage{xcolor}
\usepackage{graphicx}
\begin{document}

\copyrightyear{2025}
\copyrightclause{Copyright for this paper by its authors.
  Use permitted under Creative Commons License Attribution 4.0
  International (CC BY 4.0).}

\conference{Workshop on The Future of Human-Robot Synergy in Interactive Environments: The Role of Robots at the Workplace @ CHIWORK’25
  June 23--25, 2025, Amsterdam, NL}

\title{Designing Intent: A Multimodal Framework for Human-Robot Cooperation in Industrial Workspaces}


\author[1]{Francesco Chiossi}[%
orcid=0000-0003-2987-7634,
email=francesco.chiossi@ifi.lmu.de,
url=https://www.francesco-chiossi-hci.com/,
]
\cormark[1]

\author[1]{Julian Rasch}[%
orcid=0000-0002-9981-6952,
email=julian.rasch@ifi.lmu.de,
url=https://julian-rasch.com/,
]

\author[2]{Robin Welsch}[%
orcid=0000-0002-7255-7890,
email=robin.welsch@aalto.fi,
url=https://human-ai-interaction.com/
]

\author[1]{Albrecht Schmidt}[%
orcid=0000-0003-3890-1990,
email=albrecht.schmidt@um.ifi.lmu.de,
url=https://uni.ubicomp.net/as/,
]

\author[3]{Florian Michahelles}[%
orcid=0000-0003-1486-0688,
email=florian.michahelles@tuwien.ac.at,
url=https://informatics.tuwien.ac.at/people/florian-michahelles,
]

\address[1]{LMU Munich, Munich, Germany}
\address[2]{Aalto University, Espoo, Finland}
\address[3]{TU Wien, Visual Computing \& Human-centered Technology, Vienna, Austria}

\cortext[1]{Corresponding author.}
\begin{abstract}
As robots enter collaborative workspaces, ensuring mutual understanding between human workers and robotic systems becomes a prerequisite for trust, safety, and efficiency. In this position paper, we draw on the cooperation scenario of the AIMotive project—in which a human and a cobot jointly perform assembly tasks—to argue for a structured approach to intent communication. Building on the Situation Awareness-based Agent Transparency (SAT) framework and the notion of task abstraction levels, we propose a multidimensional design space that maps intent content (SAT1–3), planning horizon (operational to strategic), and modality (visual, auditory, haptic). We illustrate how this space can guide the design of multimodal communication strategies tailored to dynamic collaborative work contexts. With this paper, we lay the conceptual foundation for a future design toolkit aimed at supporting transparent human-robot interaction in the workplace. We highlight key open questions and design challenges, and propose a shared agenda for multimodal, adaptive, and trustworthy robotic collaboration in hybrid work environments.
\end{abstract}


\maketitle

\begin{figure*}[ht]
  \centering
  \includegraphics[width=\linewidth]{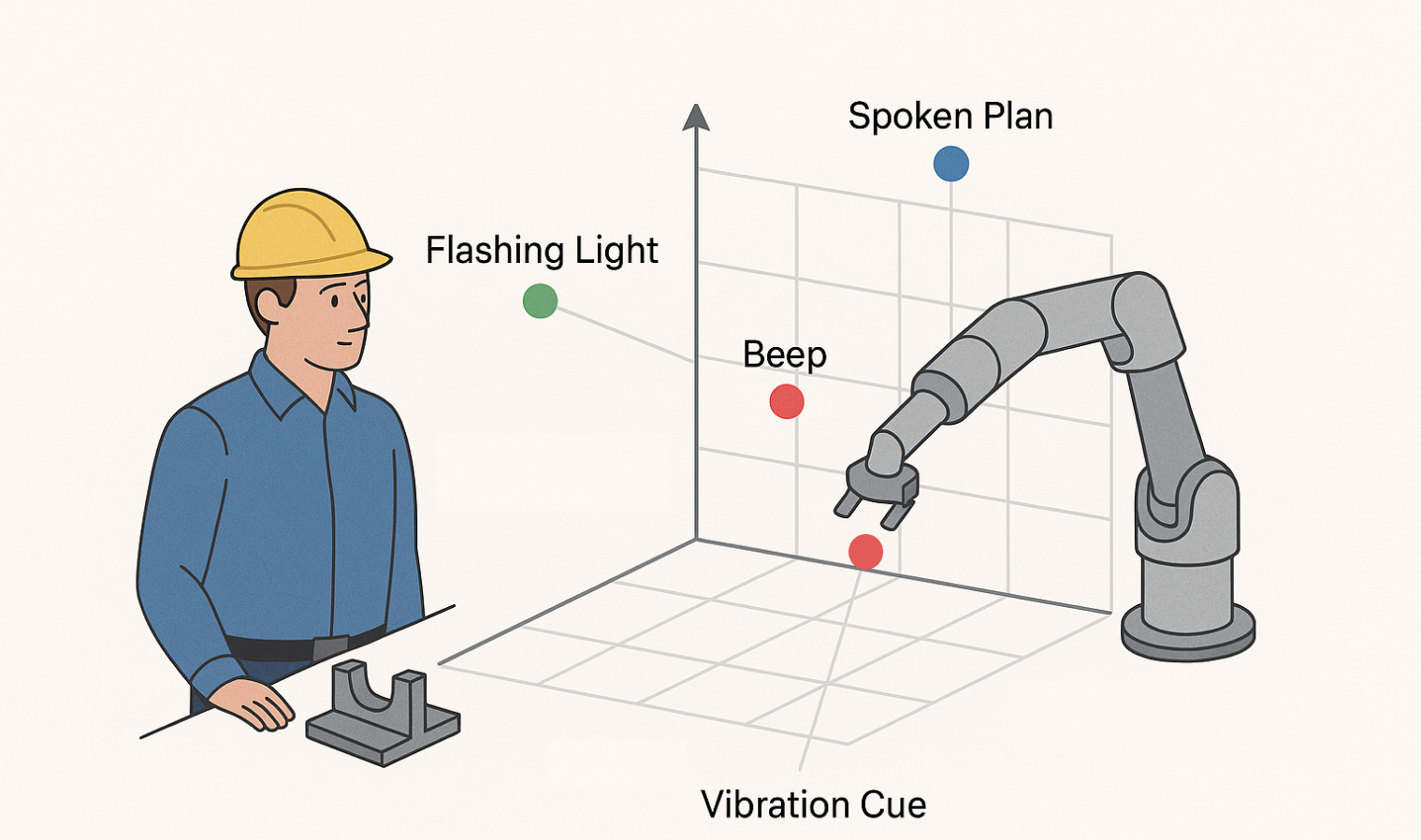} 
  \caption{In a dynamic factory workspace, a human worker and a robotic arm collaborate through a shared language of cues. The robot communicates its next move via a \textcolor{OliveGreen}{\textbf{flashing light}} (\textcolor{MidnightBlue}{\textbf{Visual}} -- \textcolor{BrickRed}{\textbf{Operational}} -- \textcolor{OliveGreen}{\textbf{SAT1}}), signals a task switch with a \textcolor{Red}{\textbf{beep}} (\textcolor{MidnightBlue}{\textbf{Auditory}} -- \textcolor{BrickRed}{\textbf{Tactical}} -- \textcolor{OliveGreen}{\textbf{SAT2}}), announces its long-term goal through \textcolor{Blue}{\textbf{speech}} (\textcolor{MidnightBlue}{\textbf{Auditory}} -- \textcolor{BrickRed}{\textbf{Strategic}} -- \textcolor{OliveGreen}{\textbf{SAT3}}), and syncs actions with the human via a subtle \textcolor{Purple}{\textbf{vibration cue}} (\textcolor{MidnightBlue}{\textbf{Haptic}} -- \textcolor{BrickRed}{\textbf{Tactical}} -- \textcolor{OliveGreen}{\textbf{SAT2}}). Each signal is carefully mapped in a three-dimensional design space—capturing what is being communicated, how, and at what level of abstraction—to ensure transparency, coordination, and trust. Image generated using ChatGPT 4o.
  }
  \label{fig:teaser}
\end{figure*}

\section{Introduction}


The increasing deployment of collaborative robots (cobots) in industrial environments is fundamentally reshaping the nature of work in manufacturing, logistics, and beyond. As humans and autonomous systems are brought into ever-closer cooperation, the ability for workers to understand, predict, and coordinate with robotic partners becomes a central requirement for both productivity and safety~\cite{, Kassem2022}. However, despite advances in robotics and artificial intelligence, a persistent challenge remains: \emph{how can cobots effectively communicate their intentions, what they are about to do, why, and how, to their human collaborators in ways that are both understandable and actionable?} This challenge is particularly acute in dynamic, safety-critical settings, where misalignment or misunderstanding can have severe consequences~\cite{Lee2004, Chen2014}.


Researchers have long recognized the importance of system transparency, predictability, and trust in human-robot interaction (HRI) and have explored a range of intent communication strategies across domains such as automated driving, smart factories, and service robotics~\cite{Kassem2022, Dey2020, Mayer2016, Mahadevan2018, Breazeal2003}. Approaches include visual cues (e.g., lights, displays, path projections), auditory alerts, and haptic feedback, as well as more implicit social signals like gaze and proxemics~\cite{Wintersberger2019, Moon2014, Walters2009}. Yet, much of this research remains fragmented, often focusing on single modalities, isolated scenarios, or specific human-robot relationships. As a result, generalizable principles for intent communication remain elusive, and the field lacks a systematic framework for comparing and designing multimodal communication strategies that work across diverse workplace scenarios~\cite{Hancock2011}. Addressing this gap is crucial, as effective intent communication is key to fostering trust, safety, and acceptance of cobots in the workplace~\cite{Endsley2017}.


In response, this paper proposes a multidimensional design space for intent communication in human robot cooperation, grounded in the conceptual and empirical foundations of the AIMotive project. Our framework systematically combines three relevant dimensions: (1) the content of communication as described by the Situation Awareness-based Agent Transparency (SAT) framework~\cite{Endsley2017}, (2) the level of task abstraction (operational, tactical, strategic), and (3) the modality of communication (visual, auditory, haptic, or multimodal). This approach enables us to map out how, what, and when intent should be communicated, tailored to different human-robot relationships and workplace contexts. The novelty of our work lies in this integrative, scenario-driven perspective, which aims to bridge the gap between fragmented prior studies and provides practical guidance for the design of transparent, trustworthy, and effective cobot systems~\cite{ Mayer2016}.

Our framework reveals that intent communication must be both multimodal and context-sensitive: the optimal combination of information content, abstraction level, and modality varies depending on the specific cooperation scenario, the criticality of the task, and the user’s role (e.g., bystander, collaborator, shared control). We identify key design opportunities and gaps, such as the need for anticipatory but non-overloading cues, and for communication strategies that generalize across domains like healthcare, logistics, and telepresence. Ultimately, we argue that a structured design space and toolkit of reusable ``intent chunks'' will be essential for building the next generation of workplace robots that are not only functionally capable but also transparent, predictable, and trustworthy partners for human workers~\cite{Kassem2022, Wintersberger2019}. We invite the HCI community to join us in validating, extending, and applying this framework to advance the future of human-robot synergy in interactive environments.

\section{Human-Cobot Cooperation in Industrial Workspaces}

The integration of collaborative robots (cobots) into industrial settings is rapidly transforming traditional work environments into dynamic, hybrid spaces where humans and autonomous systems work side by side~\cite{Mayer2016}. A representative scenario, as investigated in the AIMotive project, involves a human worker and a cobot jointly assembling parts on a production line. In such a setting, both agents must coordinate their actions, share goals, and flexibly adapt to each other's behaviors. For example, a human may align and hold a component while the cobot drills or fastens it, requiring precise turn-taking and mutual awareness~\cite{Kassem2022}.

This scenario is inspired by real-world applications in smart factories, automotive assembly, and logistics hubs, where the demand for efficient, safe, and adaptive collaboration is driving the adoption of robotic assistants~\cite{Mayer2016}. In these environments, the dynamics of human-cobot cooperation are characterized by shared objectives, distributed subtasks, and frequent safety-critical moments. Turn-taking, such as alternating between human and robot actions during assembly, requires clear communication of intent to prevent errors or accidents. Moreover, the ability to anticipate the cobot's next move is essential for maintaining workflow continuity and ensuring operator safety.

The relevance of this scenario to the HCI community is clear: it exemplifies the emergence of new forms of hybrid work, where collaboration between humans and intelligent systems is not only possible but necessary for achieving higher productivity and safety standards. As workspaces become more interactive and adaptive, understanding and designing for effective intent communication will be crucial for enabling seamless human-robot synergy~\cite{, Kassem2022}.

\section{Conceptual Foundations}

Designing effective intent communication for human-robot collaboration requires a careful consideration of what information to convey, how to present it, and when to deliver it. Broader literature converges on three foundational concepts that structure this design space: the Situation Awareness-based Agent Transparency (SAT) framework, levels of task abstraction, and the use of multimodal communication strategies~\cite{Endsley2017,Kassem2022}.

\subsection{The SAT Framework: Transparency in Perception, Reasoning, and Projection}

The SAT framework~\cite{Endsley2017} provides a systematic approach to understanding and designing transparency in autonomous systems. It distinguishes three levels of information that can be communicated to users:
\begin{itemize}
    \item \textbf{SAT1 -- Perception:} Communicates what the system currently perceives or intends, such as its present goals, environment, or status indicators.
    \item \textbf{SAT2 -- Reasoning:} Makes visible the system's internal processing or decision-making, for instance, by showing which objects are being considered in path planning or which factors influence its next action.
    \item \textbf{SAT3 -- Projection:} Provides estimates of future system states or actions, such as predicted trajectories, upcoming tasks, or anticipated outcomes.
\end{itemize}

By addressing all three SAT levels, intent communication extends beyond immediate actions to support a complete framework of \emph{why} and \emph{how} the system behaves as it does, which is critical for fostering calibrated trust and effective collaboration~\cite{Endsley2017}.

\subsection{Task Abstraction: Operational, Tactical, and Strategic Levels}

Another key dimension is the \emph{level of task abstraction}, which refers to the time horizon and granularity of the information communicated~\cite{,Chen2014}. 
\begin{itemize}
    \item \textbf{Operational level} answers ``what'' the system is doing right now, focusing on immediate actions such as moving an arm to pick up a part or activating a safety mechanism.
    \item \textbf{Tactical level} addresses ``how'' the system plans to achieve its next goal, such as selecting the order of assembly steps or choosing a path to avoid obstacles in the workspace.
    \item \textbf{Strategic level} explains ``why'' the system is acting in a certain way, relating to long-term objectives like optimizing workflow efficiency, adapting to production changes, or ensuring overall system safety.
\end{itemize}
This stratification is crucial because different users and situations require different types of information: an operator may need operational cues to avoid collisions, while a supervisor may benefit from strategic insights to understand system priorities~\cite{Chen2014}. By aligning intent communication with the appropriate level of abstraction, designers can better support decision-making, anticipation, and trust in human-robot teams.

\subsection{Multimodal Intent Communication: Principles and Relevance}

The modality of communication- the channel through which intent is conveyed- impacts how effectively information is perceived and acted upon. We emphasize that using multiple modalities can significantly enhance the clarity, accessibility, and robustness of intent communication~\cite{Kassem2022}. The principal modalities include:
\begin{itemize}
    \item \textbf{Visual:} Light indicators, AR overlays, projected paths, and displays provide immediate and spatially precise cues, making them effective for conveying status and movement intentions~\cite{Kassem2022}.
    \item \textbf{Auditory:} Voice messages, beeps, alarms, and spatialized sound can attract attention, signal urgency, and provide information even when visual attention is occupied elsewhere~\cite{Mahadevan2018,Wintersberger2019}.
    \item \textbf{Haptic:} Vibrations delivered through floors, handles, or wearable devices offer a tactile channel for communication, which is especially valuable in noisy or visually cluttered environments, or for users with sensory impairments~\cite{Chen2014}.
    \item \textbf{Multimodal:} Combining two or more modalities-such as pairing a flashing light with a beep-can reinforce critical messages, reduce the risk of missed signals, and accommodate diverse user preferences and situational constraints~\cite{,Kassem2022}.
\end{itemize}
Recent research showed the need for systematic comparison of these modalities and their combinations across different scenarios and user roles, as the optimal strategy may vary depending on the context and the specific demands of the task~\cite{Kassem2022}. For instance, redundant multimodal signals can enhance safety in high-stakes environments \cite{zhao2024survey}, while subtle haptic cues may be preferable in settings where minimizing distraction is compulsory \cite{chiossi2022design, nukarinen2014effects}.

In summary, the integration of SAT levels, task abstraction, and multimodal communication provides a conceptual foundation for designing transparent, trustworthy, and effective intent communication in human-robot collaboration. This approach not only addresses the technical challenges of information delivery but also aligns with the broader goal of enabling seamless, user-centered cooperation in increasingly complex and interactive work environments.

\section{A Multidimensional Design Space}
In this section, we present a structured design space for intent communication in human-robot collaboration, synthesizing the conceptual foundations outlined above. This design space is intended as a theoretical and practical tool to guide the analysis, design, and evaluation of communication strategies in collaborative robotic systems. It is based on three intersecting dimensions: the \textit{level of transparency} (based on the SAT framework), the \textit{level of task abstraction}, and the \textit{modality} through which information is conveyed. Each axis reflects a critical aspect of what, when, and how intent should be communicated to support effective, trustworthy interaction between human users and robotic systems.

\subsection{Dimension 1: Transparency Level (SAT)}

The first axis of the design space is grounded in the Situation Awareness-based Agent Transparency (SAT) framework~\cite{Endsley2017}, which distinguishes three types of information a system can make available to the user: what it perceives (SAT1), how it is reasoning (SAT2), and what it projects will happen next (SAT3). These levels provide a structured way of thinking about the depth of information being communicated, from immediate perceptual cues to predictive, anticipatory explanations.

\subsection{Dimension 2: Task Abstraction Level}

The second axis concerns the temporal and conceptual abstraction of the task at hand. This dimension includes operational, tactical, and strategic levels, as commonly found in both cognitive models and robotics planning literature~\cite{Chen2014, grier2012military}. The operational level corresponds to immediate, concrete actions; the tactical level represents short-term planning and sequencing; and the strategic level reflects broader goals and adaptive reasoning. This hierarchy of abstraction is essential in human-robot teaming, where different users—such as operators, technicians, or supervisors—may require different levels of granularity in the information provided.

\subsection{Dimension 3: Communication Modality}

The third axis addresses the channel through which intent is conveyed: visual, auditory, or haptic. Each modality has distinct affordances and limitations. Visual signals are spatially precise and often ambient but require line of sight. Auditory signals can draw attention and convey urgency, even when users are visually engaged elsewhere. Haptic signals provide a private and often subtle channel for information, which is especially useful in noisy or cluttered environments. Multimodal communication—using two or more channels in parallel or sequentially—can reinforce signals, improve redundancy, or reduce cognitive load when designed appropriately~\cite{Kassem2022}.

\subsection{Applying the Design Space}

The design space is conceptualized as a three-dimensional matrix (Figure~\ref{fig:designspace}) in which any unit of intent communication—what we refer to as an \textit{intent chunk}—can be positioned according to these three axes. Each chunk is defined by the type of information it conveys (SAT level), the abstraction of the task it relates to, and the modality used to communicate it. This positioning supports both the design and analysis of human-robot interaction by clarifying the purpose, timing, and channel of communication.

To illustrate the use of this model, we draw on the cooperation scenario developed in the AI Motive project. Consider a robot that assists a human in an assembly task involving drilling and fastening. In this context, a \textcolor{green!60!black}{\textbf{Flashing Light}} on the robot may signal an imminent action, such as moving the drill into position. This cue corresponds to SAT1 (perception), operational level abstraction (immediate action), and visual modality. A more complex message, such as \textcolor{blue!70!black}{\textbf{Spoken Plan}}—for example, “I will now assemble the left panel”—reflects SAT3 (projection), strategic level abstraction, and auditory modality. Finally, a \textcolor{red!70!black}{\textbf{Vibration Cue}} delivered through a wearable device could indicate readiness for synchronized movement, aligning with SAT2, tactical level, and haptic modality.

Each of these examples occupies a unique cell in the design space, but they can also be grouped or combined into multimodal bundles to suit specific roles, tasks, or user needs. Importantly, the design space also reveals underexplored areas: for instance, haptic communication at the strategic level is rare but may hold potential in supervisory control settings, while visual-tactile combinations may enhance awareness in constrained or high-risk environments.

\subsection{Implications and Opportunities}

This design space offers a systematic foundation for the development of a future \textit{toolkit} of reusable intent chunks—standardized, validated communication strategies that can be applied across different robotic systems and work scenarios. It also supports the evaluation of existing systems by revealing imbalances, such as over-reliance on visual cues at low levels of abstraction, or missing SAT3 information for long-term collaboration. By mapping intent communication strategies across these three axes, designers and researchers can reason more clearly about how to match information delivery to user goals, task demands, and contextual constraints.

Ultimately, this model advances the goal of transparent, adaptive, and user-centered human-robot collaboration by providing a shared vocabulary and structure for designing intent communication in complex work environments.

\begin{figure}
  \centering
  \includegraphics[width=\linewidth]{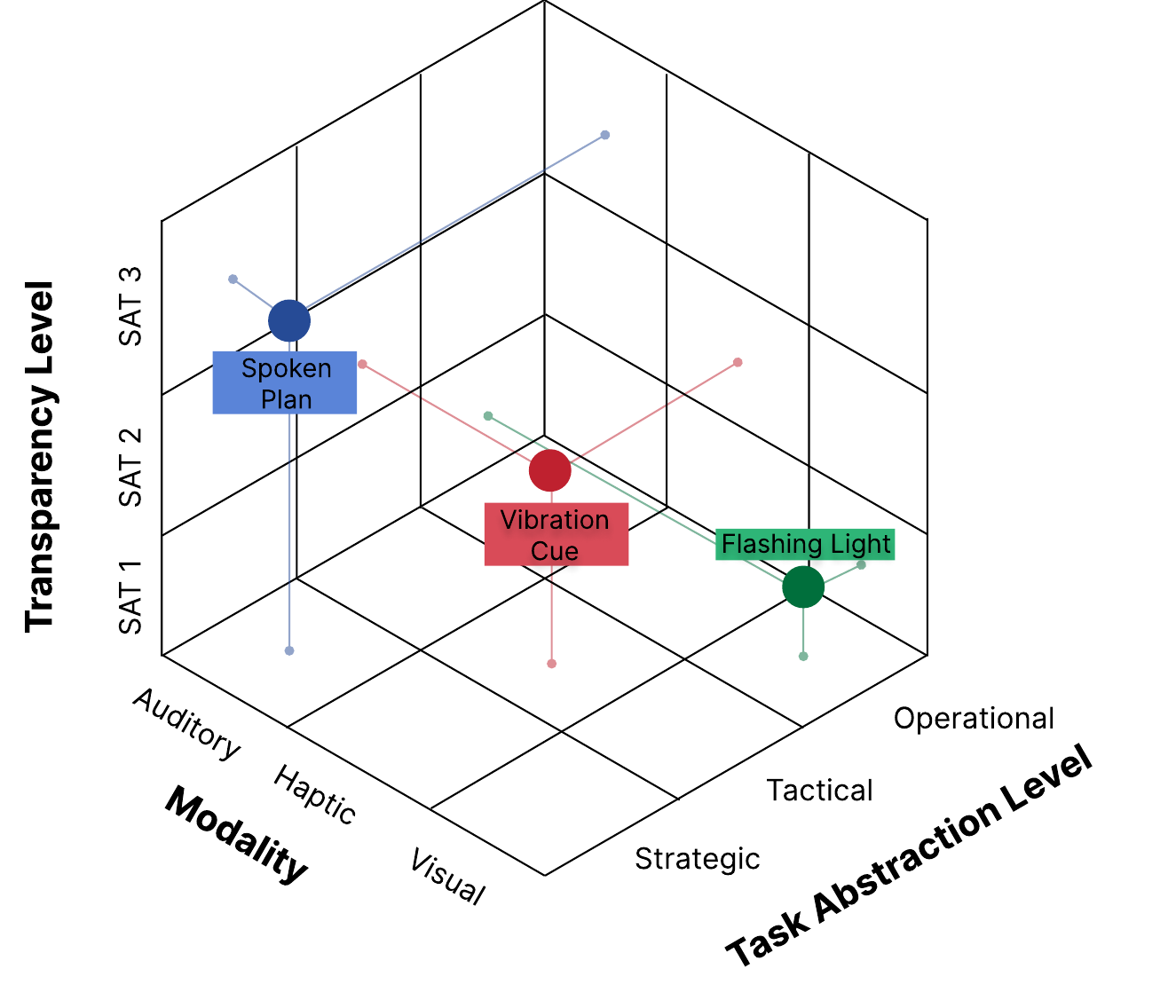}
\label{designspace}
\caption{
\textbf{Design Space for Multimodal Intent Communication in Human-Robot Cooperation.} 
This 3D framework illustrates how three key dimensions—\textbf{Task Abstraction Level} (X-axis), \textbf{Transparency Level (SAT Framework)} (Y-axis), and \textbf{Modality} (Z-axis)—interact in designing robot intent communication strategies. The X-axis ranges from \textit{Strategic} (long-term reasoning), to \textit{Tactical} (short-term planning), to \textit{Operational} (immediate actions). The Y-axis follows the Situation Awareness-based Agent Transparency (SAT) framework, from \textit{SAT1} (perception of the environment), to \textit{SAT2} (reasoning or processing), to \textit{SAT3} (projection of future actions). The Z-axis represents the communication \textbf{Modality}, spanning \textit{Visual}, \textit{Auditory}, and \textit{Haptic} channels. Each colored dot represents a specific "intent chunk"—a unit of robotic communication grounded in task timing, transparency level, and modality. 
\textbf{\textcolor{green!60!black}{\large$\bullet$}~\textcolor{green!60!black}{Flashing Light}} represents a visual signal for immediate action (Operational–SAT1–Visual), for example, a robot flashing an LED to indicate that it is about to initiate a screwing motion. 
\textbf{\textcolor{blue!70!black}{\large$\bullet$}~\textcolor{blue!70!black}{Spoken Plan}} reflects a verbal communication of long-term intention (Strategic–SAT3–Auditory), such as the robot announcing, “I will assemble the left panel next.” \textbf{\textcolor{red!70!black}{\large$\bullet$}~\textcolor{red!70!black}{Vibration Cue}} denotes a haptic signal aligned with short-term coordination (Tactical–SAT2–Haptic), for example, a vibration in a wearable device indicating that the robot is ready for synchronized movement. This framework supports systematic exploration of how robotic systems can convey intent clearly and appropriately across diverse collaborative work scenarios.
}
\label{designspace}
\end{figure}

\section{Design Considerations and Challenges}

Designing effective intent communication for human-robot collaboration is a multidimensional endeavor. It requires aligning informational content, abstraction level, and modality in a way that is not only theoretically coherent but also practically viable across diverse contexts and user needs. The design space proposed in this paper provides a structured lens through which to reason about these choices. However, its successful application hinges on addressing four central design considerations: balancing trust and cognitive load, ensuring ergonomic and inclusive modality selection, coordinating information timing, and achieving generalizability across domains.

\subsection{Trust versus Overload: Balancing Transparency with Cognitive Demand}

A primary function of intent communication is to foster calibrated trust—ensuring that users understand what the system is doing, why it is doing it, and what it will do next~\cite{Endsley2017}. Mapping intent chunks across SAT levels supports this goal by varying the depth of transparency. However, each addition of information also adds cognitive demand. Poorly timed or excessive communication, especially at higher SAT levels (e.g., strategic projections), risks overloading the user, leading to distraction, annoyance, or even distrust~\cite{Wintersberger2019}.

The design space helps address this by enabling selective emphasis: for instance, prioritizing SAT1 content during high-tempo operational tasks, and introducing SAT3 insights only when cognitive resources allow. Designers must balance the \textbf{density} and \textbf{timing} of intent chunks so that transparency enhances—rather than impairs—performance. This balance is especially critical in safety-sensitive contexts, where attentional resources are limited and the stakes of miscommunication are high.

\subsection{Ergonomics and Accessibility: Modality Selection in Context}

The choice of modality is not neutral; it directly affects the inclusiveness, usability, and effectiveness of intent communication~\cite{Kassem2022,Chen2014}. Our design space foregrounds modality as a dimension on par with informational content and task abstraction, allowing designers to explicitly consider the fit between communication channel and user context. 

Visual signals, for example, may be ideal in static, well-lit environments but inadequate in mobile or visually demanding settings \cite{chiossi2023can, chiossi2024understanding}. Auditory signals can overcome visual clutter but are easily masked in noisy environments \cite{corcoran2017sensing}. Haptic cues—such as wearable vibrations—offer a promising alternative in cases of sensory overload or for users with visual or auditory impairments~\cite{ chiossi2020notification}. 

By positioning intent chunks along the modality axis, designers can explore and evaluate alternatives that are not only efficient, but also ergonomically and socially inclusive. Furthermore, multimodal combinations—when designed deliberately rather than redundantly—can help mitigate single-channel limitations without increasing user burden.

\subsection{Temporal Coordination: Aligning Timing with Task Flow}

Effective intent communication is not merely about what is said, but when it is said. Poor timing undermines otherwise useful signals. Delivering information too early may decouple it from relevance; delivering it too late can impede reaction or coordination~\cite{Wintersberger2019}. The vertical SAT dimension of our design space (perception → projection) interacts with this temporal challenge, as different levels of information imply different timing windows.

For example, SAT1 cues (e.g., “the robot is about to move”) must be delivered with sufficient lead time to avoid surprise, whereas SAT3 messages (e.g., “the robot intends to complete this task group”) may need to be presented when users are cognitively available to absorb broader context. Designers can use the design space not only to select \textbf{which} intent chunks to deliver, but also to plan \textbf{when} they are most helpful, scaffolded by the abstraction level and modality best suited to the pace of the task.

\subsection{Generalizability: From Scenario-Specific Design to Cross-Domain Principles}

One of the central aspirations of the AI Motive project is to develop intent communication strategies that extend beyond individual use cases. Our design space serves as a vehicle for generalization: by abstracting intent chunks in terms of their SAT level, planning horizon, and modality, we create a transferable framework that can apply across domains, from collaborative manufacturing to assistive robotics and telepresence systems~\cite{,Kassem2022}.

Scenario-driven development remains essential to ground design choices in real-world constraints. However, the structure of the design space enables comparative reflection: are certain modality–SAT combinations more robust across contexts? Are some intent chunks universally interpretable? Are certain dimensions more sensitive to domain shift than others? Answering these questions requires systematic exploration of the space, but the model presented here provides a shared vocabulary and scaffold to support that process.

In summary, the considerations outlined in this section pinpoint the value, and necessity, of a structured design framework. We considered trust, accessibility, timing, and generalizability within the design space; thus moving closer to establishing a principled and adaptable foundation for intent communication in human-robot collaboration.


\section{Conclusion}


In this paper, we introduced a multidimensional design space that links the content of intent communication (\textit{what} is conveyed, via SAT levels), its timing (\textit{when}, via task abstraction), and the communication channel used (\textit{how}, via modality). Framed through illustrative cooperation scenarios, this model enables a more systematic approach to designing intent communication strategies in human-robot interaction. We conceptualized intent as modular “chunks” that occupy distinct positions in this space, and we provided a practical framework for designing and evaluating transparent robotic behaviors.

We do not propose this space as a final solution, but as a generative scaffold for ongoing research, evaluation, and tool development. We invite the CHIWORK and HRI communities to engage with this framework by refining its structure, testing its applicability across domains, and contributing new use cases. Our long-term goal is to co-develop a shared toolkit that supports transparent, trustworthy, and adaptive collaboration between humans and robots in the future of work.

\begin{acknowledgments}
Francesco Chiossi was supported by the Deutsche Forschungsgemeinschaft (DFG, German Research Foundation) with Project ID 251654672 TRR 161.
\end{acknowledgments}



\section*{Declaration on Generative AI}
During the preparation of this work, the author(s) used OpenAI’s GPT-4 and Grammarly for grammar, generation of Figure 1, and style editing. All content was reviewed and edited by the author(s), who take full responsibility for the final publication.
\bibliography{references}


\end{document}